 \documentclass[final,3p,times]{elsarticle}

\usepackage[english]{babel}
\usepackage[latin1]{inputenc}
\usepackage{amsfonts} 
\usepackage{amsmath}
\usepackage{amssymb}
\usepackage{amsthm}
\usepackage{enumerate}
\usepackage{hyperref} 
\usepackage{graphicx} 
\usepackage{tabularx} 
\usepackage[usenames,dvipsnames]{color}
\usepackage{rotating}
\usepackage{epsfig} 
\usepackage{graphicx}
\usepackage{threeparttable}
\usepackage{multirow}
\usepackage{url}
\usepackage{booktabs} 
\usepackage{lineno}

\newcolumntype{Y}{>{\centering\arraybackslash}X}

\journal{Journal  }

\begin{document}

\begin{frontmatter}


\title{Decision tool and Sample Size Calculator for composite endpoints}



\author{Marta Bofill Roig}
\ead{marta.bofill.roig@upc.edu}
\author{Jordi Cortés Mart\'inez}
\ead{jordi.cortes-martinez@upc.edu}
\author{Guadalupe G\'omez Melis}  
\ead{lupe.gomez@upc.edu}

\address{Departament d'Estad\'{i}stica i Investigaci\'{o} Operativa, Universitat Polit\`{e}cnica  de Catalunya, Barcelona, Spain}


\end{frontmatter}



\textbf{Summary Points:}
	{\small
	\begin{itemize}
		\item This article considers the combination of two binary or two time-to-event endpoints to form the primary composite endpoint for leading a trial. 
		\item It discusses the relative efficiency of choosing a composite endpoint over one of its components in terms of: the frequencies of observing each component; the relative treatment effect of the tested therapy; and the association between both components.
		\item We highlight the very important role of the association between components in choosing the most efficient endpoint to use as primary.
		\item For better grounded future trials, we recommend trialists to always reporting the association between components of the composite endpoint 
		\item Common fallacies to note when using composite endpoints: i) composite endpoints always imply higher power; ii) treatment effect on the composite endpoint is similar to the average effects of its components; and iii) the probability of observing the primary endpoint increases significantly.  
\end{itemize}}

\medskip

\hrule
\bigskip
\hrule 

\section{INTRODUCTION} 
Choosing the primary endpoint (PE) in a randomized controlled trial (RCT) is a critical decision in the design phase. Composite endpoints are defined as the occurrence of any of the relevant events in trials with binary response and as the time from randomization to the first observed event among all components in time-to-event studies. Some examples are the composite of death or myocardial re-infarction and major adverse cardiovascular events (MACE) generally defined as a composite of cardiovascular death, myocardial infarction, stroke and target vessel revascularization. 

What are the reasons for using a CE? On the one hand, taking into account multiple outcomes altogether could provide a broader overview of an intervention's efficacy (1). On the other hand, if the event rates are too low, it could increase the study power. However, CEs also hide some drawbacks. For instance, the tested intervention should have a similar effect on all the components while none of them should be affected negative (2). Moreover, having more outcomes does not necessarily imply an increase in power: small intervention effects on one component could mask significant effects on others (3) or strong associations may reduce the joint information.

We discuss the gain in efficiency of adding endpoints to form a primary CE rather than simply using a single relevant endpoint. We provide a multitask tool named CompARE that can be freely access through \url{https://cinna.upc.edu/compare/} as an aid for a more well-informed decision. The code of the functions implemented in this web app is available in GitHub at  \url{https://github.com/CompARE-Composite/Functions}. \href{https://cinna.upc.edu/compare/}{CompARE} allows to check the efficiency of different designs with more than one outcome and provides a calculator to get the required sample size based on several CEs configurations among other features. This will be illustrated through two RCT published in NEJM and European Heart Journal, respectively (4,5). 

\noindent\fbox{%
	\parbox{\textwidth}{%
		\textbf{List of abbreviations}
		{\small
			\begin{itemize}
				\item ARE:	Asymptotic Relative Efficiency
				\item $\varepsilon_1$, $\varepsilon_2$, $\varepsilon_*$:	Relevant event, $\varepsilon_1$; Additional event, $\varepsilon_2$;  Composite event, $\varepsilon_*$, defined as $\varepsilon_1$ or $\varepsilon_2$.
				\item HR:	Hazard Ratio 
				\item PE:	Primary Endpoint
				\item OS:	Overall survival
				\item MI:	Myocardial infarction 		
		\end{itemize}}
	}%
}

\section{RELATIVE EFFICIENCY OF USING COMPOSITE ENDPOINTS} \label{Sect2}

Should we use the most relevant endpoint (such as overall survival, OS) as the PE for a two-arm RCT? Or should we base our trial on a CE that includes an additional endpoint (such as myocardial re-infarction)? Both endpoints are capable of measuring the treatment effect in the same direction, are of interest to the investigator but they are not equally efficient. 

What does efficiency mean when it refers to an endpoint that may be chosen for testing a therapy or treatment? According to the Oxford dictionary, a system is efficient when it ``achieves maximum productivity with minimum wasted effort or expense". A hypothesis testing procedure based on a given endpoint is efficient when it can be influenced by the treatment achieving maximum power for a given sample size. The efficiency is then a measure of the quality of a given test based on an endpoint, while the relative efficiency between two endpoints is determined by their efficiency ratio. Since often we can only assess the efficiency of a statistical test for large enough sample sizes, we use the asymptotic relative efficiency (ARE) as the principal comparison measure. This measure has been studied in a time-to-event framework (6) as well as in a binary setup (7) and it has been used (8) to assess, in the cardiovascular research area, the characteristics that a potential endpoint should have to be part of the CE leading the trial. 

Suppose that, two potential endpoints could satisfactorily answer the study's primary clinical question: we could use a relevant event, say $\varepsilon_1$, or the composite event ($\varepsilon_*$), say $\varepsilon_1$ or $\varepsilon_2$, where $\varepsilon_2$ is some additional endpoint. The ARE measure can be roughly interpreted as the ratio of the required sample sizes using $\varepsilon_1$ versus $\varepsilon_*$ to attain the same power for a given significance level (9) yielding the following criterion: whenever ARE $>$1, choose $\varepsilon_*$ as PE to guide the study; otherwise, use $\varepsilon_1$.  		  

To calculate the ARE we need to anticipate:  the frequencies of observing events $\varepsilon_1$ and $\varepsilon_2$ in the control group; the relative treatment effect given by either the hazard ratios (HR) --survival outcome-- or risk differences --binary outcome-- and a measure of the association between components. In very general terms, one should use the CE (ARE$>$1) if the relative effect of treatment on the additional endpoint is greater or about the same as that on the relevant endpoint. Nevertheless, other scenarios such as those for slightly smaller relative effect of treatment on the additional endpoint together with low frequencies of observing the relevant endpoint, might yield ARE$>$1, thus, making the composite endpoint more efficient. Furthermore, as we discuss next, the association between $\varepsilon_1$ and $\varepsilon_2$ plays a major role when computing the ARE. The ability to make a well-informed decision can be aided by a simple and friendly tool, such as \href{https://cinna.upc.edu/compare/}{CompARE}, which allows calculating the ARE in various clinical trial scenarios.


\section{ASSOCIATION BETWEEN THE COMPONENTS OF THE COMPOSITE ENDPOINT} \label{Sect3} 

The association between $\varepsilon_1$ and $\varepsilon_2$ plays a major role when computing the ARE. Two endpoints are positively associated when a large value of one implies a large value of the other, and vice versa. 

The association between two binary outcomes can be measured by means of Pearson's correlation, although the conditional probability of one, given the other, might furnish a more natural and interpretable understanding of association. When we wonder about the percentage of patients ends up suffering a stroke after having a myocardial infarction (MI), the interpretation is quite clear in terms of conditional probability. The positive and linear relationship between the correlation coefficient and the conditional probability allows to use the estimated conditional probability to guess the correlation. For instance, if the frequency of having a MI and suffering a stroke are roughly equal and very low (say, less than 5\%), the correlation between them is approximately equal to the proportion of patients having a MI among those that had a stroke. 

\medskip

\noindent\fbox{%
	\parbox{\textwidth}{%
		\textbf{ARE insights}
		{\small
			\begin{itemize}
				\item The clinical question could be satisfactorily answered by events $\varepsilon_1$ and $\varepsilon_*$ = $\varepsilon_1$ or $\varepsilon_2$, where $\varepsilon_2$ is some additional endpoint.	
				\item ARE quantifies how much more efficient the composite endpoint $\varepsilon_*$ is over $\varepsilon_1$ as the primary endpoint of the study:
				\begin{itemize}
					\item If ARE $>$1, choose $\varepsilon_*$ to guide the study; 
					\item If ARE $<$1, base the study on $\varepsilon_1$.
				\end{itemize}  
				\item To compute ARE, we need to anticipate:
				\begin{itemize}
					\item frequencies of observing   $\varepsilon_1$ and $\varepsilon_2$ in the control group,
					\item relative treatment effect given by the hazard ratios (in a survival trial) or risk differences (in a binary trial). In the survival setting, additional assumptions should be made. 
					\item a measure of the association between $\varepsilon_1$ and  $\varepsilon_2$.
				\end{itemize}			
		\end{itemize}}
	}%
}

\medskip

The association between the time-to-event outcomes for $\varepsilon_1$ and $\varepsilon_2$ is a little bit ``tricky" in the presence of competing risks, when one outcome could prevent the observation of the other. The ARE method is based on a joint copula model built on the cause specific hazards for times to $\varepsilon_1$ and $\varepsilon_2$  and  on  Spearman's rho correlation between these times. We strongly remark here that the anticipation of Spearman's rho is only done for the purpose to obtain precise sample size calculations and not assumed to be used or computed in the analysis phase when it makes no sense to get the association between times to $\varepsilon_1$ and $\varepsilon_2$ when one of them is death. 

It has been shown (6,7) that the efficiency of using a CE rather than using its more relevant component decreases as the composite components become more associated; or, in other words, the more correlated the components are, the larger the sample size needed for the trial (9) and using the CE might not be worthwhile.

The association between the components of the composite endpoint has a significant impact both on the probability of observing the composite endpoint and on the efficiency of the composite endpoint as a primary endpoint. Therefore, it is of utmost importance to consider how the components of the composite endpoint are associated, quantify this association and report them in all documents and summaries derived from the trials.

\section{CASE STUDIES USING COMPARE} \label{Sect4}

We illustrate the PE decision using two RCTs and with the aid of \href{https://cinna.upc.edu/compare/}{CompARE}, which is useful for different purposes as shown in Figure 1. \href{https://cinna.upc.edu/compare/}{CompARE} is a comprehensive and freely available web-tool intended to provide guidance on how to deal with composite endpoints in the planning stage of a randomized controlled trial. \href{https://cinna.upc.edu/compare/}{CompARE} can also be used to:   

\begin{enumerate}
	\item Choose the best primary endpoint to lead the trial. \href{https://cinna.upc.edu/compare/}{CompARE} computes the ARE to quantify the gain in the efficiency of using --as the primary endpoint-- a composite endpoint over one of its components.
	\item Specify the treatment effect for the composite endpoint based on the marginal information of the composite components and to study the performance of the composite parameters according to these. In a survival trial, it can also evaluate the proportional hazards assumption for the composite endpoint.
	\item Determine the sample size for different situations, such as when the association between composite components is unknown or when the hazards are not proportional. 
	\item Calculate and interpret the different association measures among the composite components.
\end{enumerate} 

The code to reproduce the following two case studies is available in \href{https://github.com/CompARE-Composite/Functions}{GitHub}.

\begin{figure}[h!]
	\centering
	\includegraphics[width=0.7\linewidth]{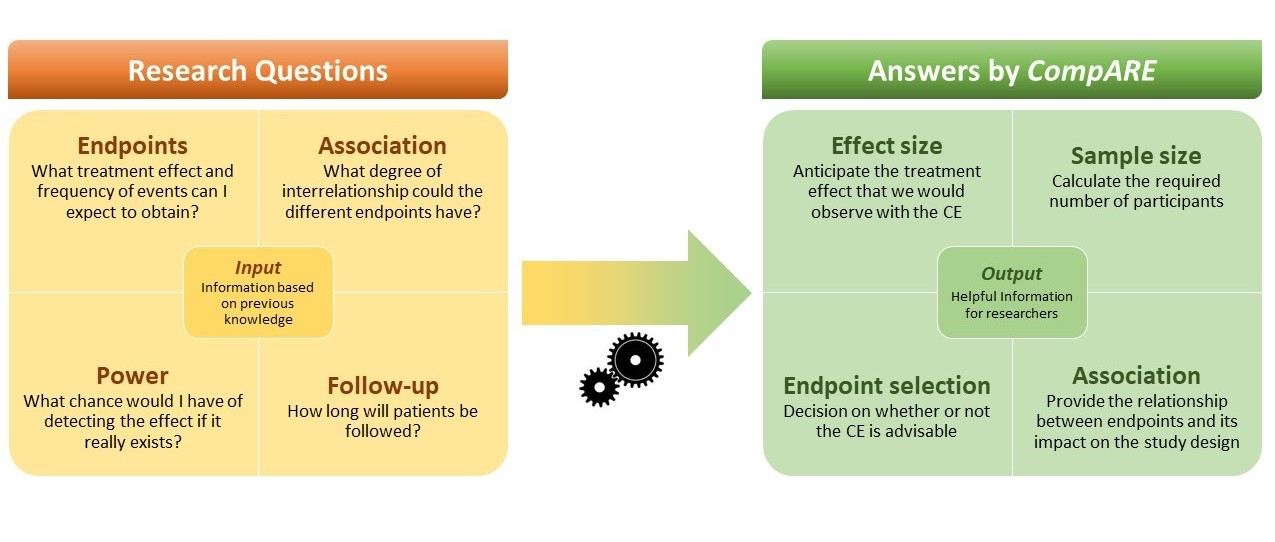}
	\caption{\href{https://cinna.upc.edu/compare/}{CompARE} scheme with the inputs to be provided by the researcher and the outputs returned by the application.}
	\label{fig:fig1}
\end{figure}

\subsection{TUXEDO trial:  binary endpoints}

The TUXEDO study was a RCT aimed at comparing paclitaxel-eluting stents (control group) with everolimus-eluting stents (intervention group) (4). The PE was target-vessel failure ($\varepsilon_*$), defined as a composite of cardiac death, target vessel MI or ischemia-driven target-vessel revascularization at 1-year of follow-up. The secondary endpoints were, among others, the ischemia-driven target-lesion revascularization ($\varepsilon_1$, revascularization for short) and the endpoint of cardiac death or target-vessel MI ($\varepsilon_2$, death or MI for short). The TUXEDO protocol was based on the results of the SPIRIT IV trial (10), which faced similar research questions and considered the same PE and that we will use as reference values for illustrative purposes. The association between the different outcomes was not reported, nor was the number of patients who had more than one event.

Would it have been preferable to consider revascularization ($\varepsilon_1$) instead of the CE of target-vessel failure ($\varepsilon_*$) as PE?  We discuss it considering that the frequency of $\varepsilon_1$ and  $\varepsilon_2$ in the control group were 5.9\% and 3.2\%, respectively. Everolimus-eluting stents as compared with paclitaxel-eluting stents resulted in a 1.96 percentage-point absolute reduction in $\varepsilon_1$ and a 0.98 percentage-point absolute reduction in $\varepsilon_2$. 

\begin{table}[h!]
	\begin{tabularx}{\textwidth}{ccccccc}
		\hline 
		Association & Correlation & Conditional   & Conditional    & Probability of   & Percentage-point  & Total   \\ 
		&  &  probability &  probability &  observing $\varepsilon_*$ &  absolute reduction  &   Sample Size \\ 
		&  &  $\varepsilon_1$ given $\varepsilon_2$ &  $\varepsilon_2$ given $\varepsilon_1$ &    &  $\varepsilon_*$ &     \\ 
		\hline 
		Weak & 0.1 &	0.19 &	0.1 &	0.08&	2.7&	2,187 \\ 
		Moderate &	0.4 &	0.58 &	0.31&	0.07&	2.3	& 2,561 \\ 
		Strong &	0.7	&0.97&	0.52&	0.06&	2.0&	3,076 \\ 
		\hline 
	\end{tabularx}
	\caption{Probability of target-vessel failure ($\varepsilon_*$) and total sample size to achieve a power of 80\% at a significance level of 0.05 in a design with $\varepsilon_*$ as PE based on the SPIRIT IV trial values and for different degrees of association between $\varepsilon_1$ and $\varepsilon_2$ }
\end{table}   

Table 1 presents three scenarios according to weak, moderate, strong degrees of association between revascularization and death or MI. The strength of association is provided both in terms of the correlation coefficient (0.1, 0.4, 0.7) and as the conditional probability of $\varepsilon_2$  given a $\varepsilon_1$. For each of these scenarios and based on the values provided by the SPIRIT IV trial, the probability of observing $\varepsilon_*$ in the control group ranges between 6\% and 8\%; and the expected absolute reduction for $\varepsilon_*$ takes values between 2.3 and 2.9. Observe that both the probability and the expected effect for $\varepsilon_*$ decrease as the association increases, and this pattern is always found (11,12). We have also computed the required total sample size for a trial with $\varepsilon_*$ as PE. Notice that, as the association between the composite components gets larger, the sample size is increasingly higher.

Figure 2 shows that the efficiency of using $\varepsilon_*$ as PE rather than $\varepsilon_1$ decreases as the components become more correlated. Indeed, revascularization is preferred over target-vessel failure when $\varepsilon_1$ and $\varepsilon_2$ are strongly associated. To study the extent to which adding $\varepsilon_2$ affects the PE choice, we consider several effect values for $\varepsilon_2$: for a given correlation, the ARE increases as the anticipated treatment effect in $\varepsilon_2$ gets larger.

\begin{figure}[h!]
	\centering
	\includegraphics[width=0.6\linewidth]{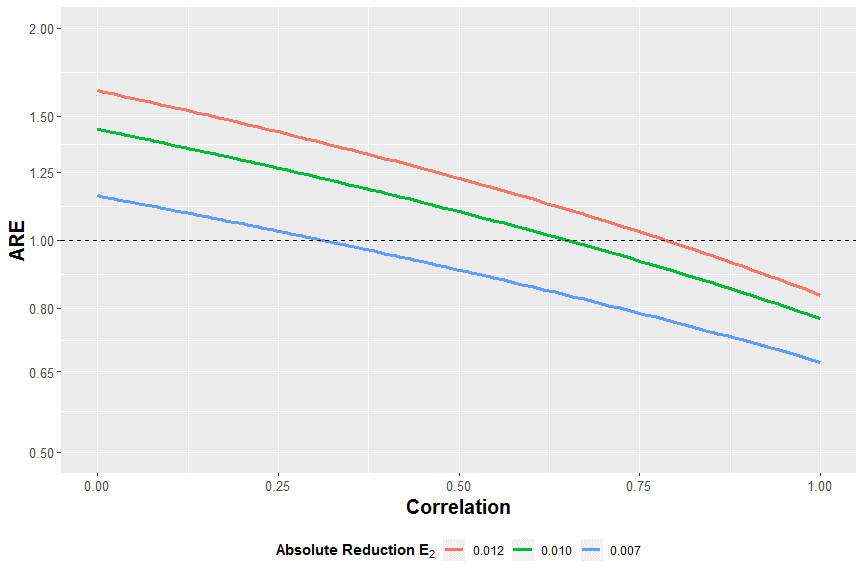}
	\caption{ARE values depending on the correlation (horizontal axis) and the treatment effect of the additional endpoint (colored lines). }
	\label{fig:fig2}
\end{figure}

\subsection{OASIS-6 trial: time-to-event endpoints}

Odlgren et al. (5) tested fondaparinux treatment vs. placebo or unfractionated heparin, in a pre-specified subgroup of 2,867 patients not receiving reperfusion treatment in the OASIS-6 trial. The primary outcome was the composite endpoint ($\varepsilon_*$) of death or myocardial re-infarction at 30 days. The observed frequency in the control group was 0.125 for OS ($\varepsilon_1$) and 0.037 for myocardial re-infarction ($\varepsilon_2$), and the estimated HRs were 0.83 for $\varepsilon_1$ and 0.66 for $\varepsilon_2$. 

If a new similar study was planned, should we use the composite endpoint ($\varepsilon_*$) to lead the trial or should we stick to OS as the PE? Figure 3 shows how \href{https://cinna.upc.edu/compare/}{CompARE} can help to discuss the efficiency of using death or myocardial re-infarction as PE (see also Figure 4), to compare sample sizes and to illustrate effect sizes for several scenarios with different HRs (between 0.65 and 0.90) for $\varepsilon_2$ and several degrees of association.

\begin{figure}[h!]
	\centering
	\includegraphics[width=0.7\linewidth]{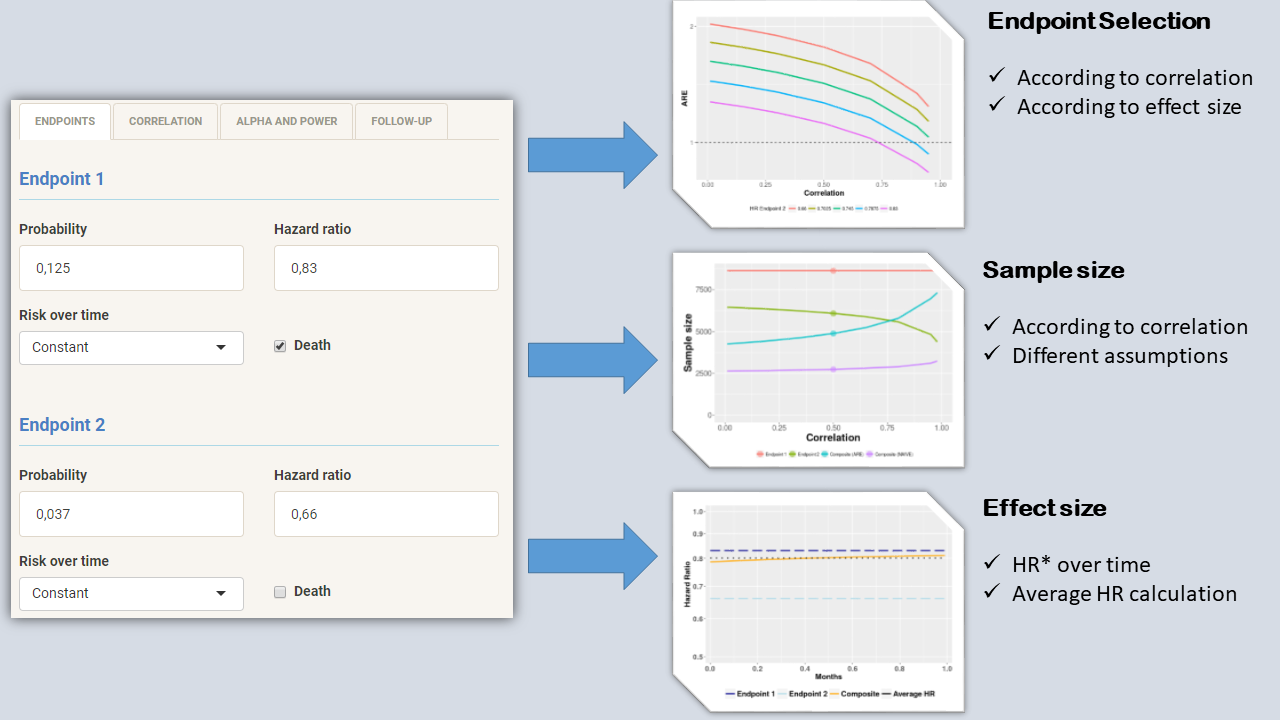}
	\caption{Screenshots of \href{https://cinna.upc.edu/compare/}{CompARE} using the information from the OASIS-6 Trial. }
	\label{fig:fig3}
\end{figure}

Figure 4 (self-explained) shows the influence of correlation on the ARE value and, consequently, can be used as a criterion to choose between OS or the CE formed by death ($\varepsilon_1$) and Myocardial re-infarction ($\varepsilon_2$). The ARE values (vertical axis) depend on the Spearman correlation (horizontal axis) and the treatment effect of the additional endpoint (colored lines). Cause-specific hazard function for the time to Myocardial re-infarction ($\varepsilon_2$) and the time to death ($\varepsilon_1$) are assumed constant during the follow-up.

When the PE is based on the time from randomization to $\varepsilon_*$, the behavior of the corresponding cause-specific hazards has to be as well anticipated. Table 2 presents the ARE and the required sample size under five selected scenarios, assuming constant, increasing or decreasing hazards for each outcome and strong correlation between them. ARE is greater than 1 in all scenarios and hence $\varepsilon_*$ is always recommended; however sample sizes vary a lot -- in a range from 2,682 to 3,381 -- exemplifying the impact of the hazard shape over time. 

\begin{figure}[h!]
	\centering
	\includegraphics[width=0.7\linewidth]{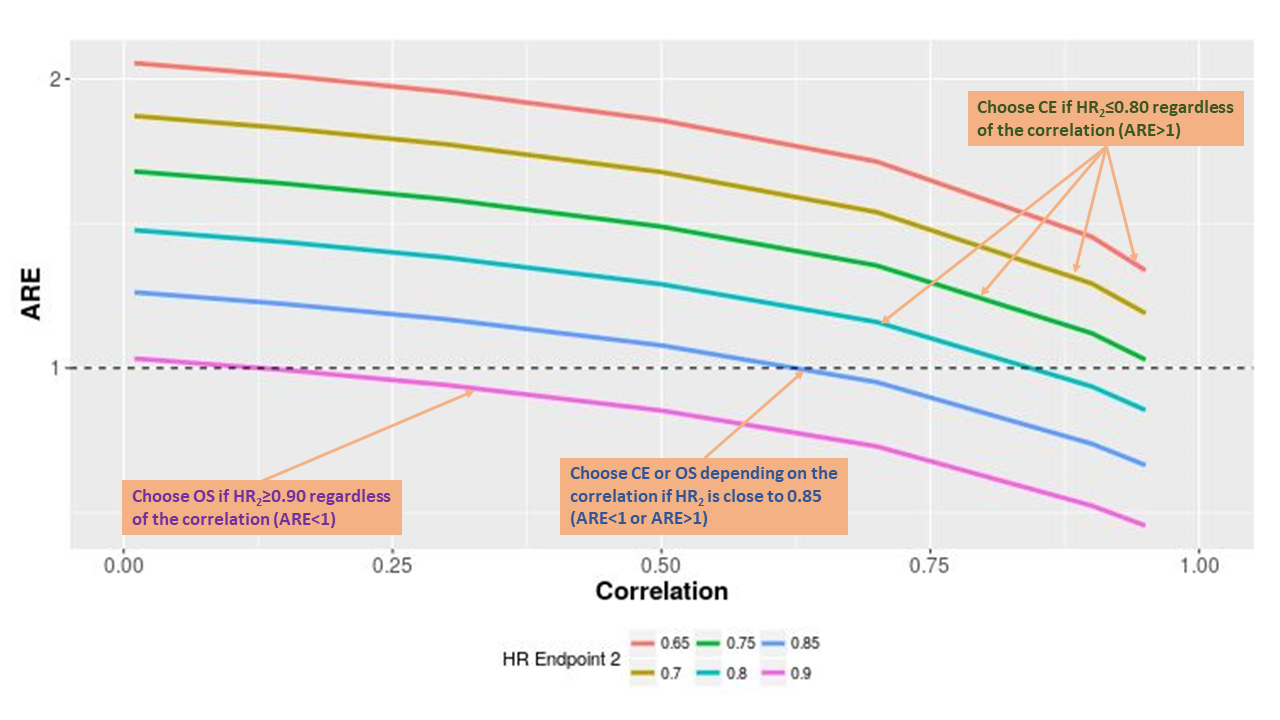}
	\caption{ARE values depending on the correlation (horizontal axis) and the treatment effect of the additional endpoint (colored lines).  }
	\label{fig:fig4}
\end{figure}

\begin{table}[h!] 
	\begin{tabularx}{\textwidth}{@{}YYYY@{}}
		\hline 
		Hazard of death & Hazard of  & ARE & Total  \\ 
		& myocardial re-infarction &  & Sample Size \\ 
		\hline 
		Increasing  $\uparrow$ &	Decreasing $\downarrow$&	1.84&	3,381  \\ 
		Constant  $\leftrightarrow$	&Decreasing $\downarrow$&	1.90&	3,278 \\
		Constant  $\leftrightarrow$&	Constant  $\leftrightarrow$&	2.02&	3,084 \\
		Constant  $\leftrightarrow$&	Increasing  $\uparrow$&	2.18&	2,857 \\
		Decreasing $\downarrow$ &	Increasing  $\uparrow$ &	2.32 &	2,682 \\
		\hline 
	\end{tabularx} 
	\caption{ARE and the required sample size to achieve an 80\% power at a significance level of 0.05 for detecting differences in the CE, depending on the cause-specific hazard behavior over time for each component: constant (exponential distribution), increasing (Weibull distribution with shape parameter equals 2) or decreasing (Weibull with shape parameter equals 0.5). Other assumed parameters: i) frequencies of 0.125 and 0.05 for death and myocardial re-infarction in the control arm; ii) HRs of 0.83 and 0.66 for OS and myocardial re-infarction; and iii) Spearman's rho correlation between components equal to 0.7, considering Gumbel copula. Sample size calculations are based on the Freedman formula.}
\end{table}

\section{DISCUSSION} 

Composite endpoints are commonly used in clinical trials and observational studies. However, contrary to popular belief, composite endpoints only barely increase total event-probability and do not necessarily yield higher power. 

The design of a randomized controlled trial involving a composite endpoint needs a careful specification of the expected rates, the treatment effects and the correlation between the components. While the frequencies and relative effects of the primary and secondary endpoints in most of the published clinical trials can often be derived from previous studies, the degree of association between them are seldom disclosed.  \href{https://cinna.upc.edu/compare/}{CompARE} can be used to explore different scenarios and for each one calculate the treatment effect for the CE and the needed sample size.

Because the association between the composite components plays an important role in the overall expected frequency of the CE, but also in the statistical power of the study, this information is critical in designing a trial, it has a large impact on the design efficiency, notably on the PE choice and, hence, on the trial's sample size. For example, the SAMPL guidelines (13) enumerate the items to be reported for each type of analysis, but currently do not make any reference to this point.  We urge trialists to report, as accurately as possible, all the primary and secondary parameter information as well as a measure of the association between the endpoints considered in the trial.


\section*{Acknowledgements}

This work is partially supported through grant MTM2015-64465-C2-1-R (MINECO/FEDER, UE) from the Secretaría de Estado de Investigación, Desarrollo e Innovación del Ministerio de Economía y Competitividad (Spain) and through grant 2017 SGR 622 (GRBIO) from the Departament d'Economia i Coneixement de la Generalitat de Catalunya (Spain). Marta Bofill Roig acknowledges financial support from the Spanish Ministry of Economy and Competitiveness, through the María de Maeztu Programme for Units of Excellence in R\&D (MDM-2014-0445).


\section*{REFERENCES}

\noindent
1. Anker SD, Schroeder S, Atar D, Bax JJ, Ceconi C, Cowie MR et al. Traditional and new composite endpoints in heart failure clinical trials: Facilitating comprehensive efficacy assessments and improving trial efficiency. Eur J Heart Fail. 2016;18(5):482-9. \\
2. European Medicines Agency. Guideline on multiplicity issues in clinical trials (draft). Vol. 44, Guidance. 2016.  \\
3. Prieto-Merino D, Smeeth L, Van Staa TP, Roberts I. Dangers of non-specific composite outcome measures in clinical trials. BMJ. 2013;347(November):1-6. \\
4. Kaul U, Bangalore S, Seth A, Priyadarshini A, Rajpal KA, Tejas MP et al. Paclitaxel-Eluting versus Everolimus-Eluting Coronary Stents in Diabetes. N Engl J Med. 2015;373(18):1709-19. \\
5. Oldgren J, Wallentin L, Afzal R, Bassand JP, Budaj A, Chrolavicius S et al. Effects of fondaparinux in patients with ST-segment elevation acute myocardial infarction not receiving reperfusion treatment. Eur Heart J. 2008;29:315-323. \\
6. Gómez G, Lagakos SW. Statistical considerations when using a composite endpoint for comparing treatment groups. Stat Med. 2013;32(5):719-38.
7. Bofill Roig M, Gómez Melis G. Selection of composite binary endpoints in clinical trials. Biometrical J. 2018;60(2):246--61. \\
8. Gómez G, Gómez-Mateu M,  Dafni U. Informed Choice of Composite End Points in
Cardiovascular Trials. Circ Cardiovasc Qual Outcomes. 2014;7:170-178. \\
9. Gómez G, Gómez-Mateu M. The asymptotic relative efficiency and the ratio of sample sizes when testing two different null hypotheses. SORT. 2014;38(1):73-88.  \\
10. Nikolsky E, Lansky AJ, Sudhir K, Doostzadeh J, Cutlip DE, Piana R et al. SPIRIT IV trial design: A large-scale randomized comparison of everolimus-eluting stents and paclitaxel-eluting stents in patients with coronary artery disease. Am Heart J. 2009;158(4):520-526.e2. \\
11. Bofill Roig M, Gómez Melis G. A new approach for sizing trials with composite binary endpoints using anticipated marginal values and accounting for the correlation between components. Stat Med. 2019;38(11):1935-56.  \\
12. Marsal JR, Ferreira-González I, Bertran S, et al. The use of a binary composite endpoint and sample size requirement: Influence of endpoints overlap. Am J Epidemiol. 2017;185(9):832-41. \\
13. Lang TA, Altman DG. Basic statistical reporting for articles published in clinical medical journals: the SAMPL Guidelines. In: Science Editors' Handbook [Internet]. 2013.p.29--32.

%
%

\clearpage 

\end{document}